\documentclass[reprint,pra, superscriptaddress,twocolumn,amsmath,amssymb]{revtex4-2} 

\usepackage{siunitx} 
\usepackage{dcolumn}
\usepackage[caption=false]{subfig}
\usepackage{graphicx}

\usepackage{bm}

\usepackage{listings}
\begin{document}

\preprint{APS/123-QED}

\title{Analysis of Proton Bunch Parameters in the AWAKE Experiment}

\author{V.~Hafych}
\affiliation{Max Planck Institute for Physics, Munich, Germany}
\author{A.~Caldwell}
\affiliation{Max Planck Institute for Physics, Munich, Germany}

\author{R.~Agnello}
\affiliation{Ecole Polytechnique Federale de Lausanne (EPFL), Swiss Plasma Center (SPC), Lausanne, Switzerland}
\author{C.C.~Ahdida}
\affiliation{CERN, Geneva, Switzerland}
\author{M.~Aladi}
\affiliation{Wigner Research Center for Physics, Budapest, Hungary}
\author{M.C.~Amoedo Goncalves}
\affiliation{CERN, Geneva, Switzerland}
\author{Y.~Andrebe}
\affiliation{Ecole Polytechnique Federale de Lausanne (EPFL), Swiss Plasma Center (SPC), Lausanne, Switzerland}
\author{O.~Apsimon}
\affiliation{University of Liverpool, Liverpool, UK}
\affiliation{Cockcroft Institute, Daresbury, UK}
\author{R.~Apsimon}
\affiliation{Cockcroft Institute, Daresbury, UK} 
\affiliation{Lancaster University, Lancaster, UK}
\author{A.-M.~Bachmann}
\affiliation{Max Planck Institute for Physics, Munich, Germany}
\author{M.A.~Baistrukov}
\affiliation{Budker Institute of Nuclear Physics SB RAS, Novosibirsk, Russia}
\affiliation{Novosibirsk State University, Novosibirsk, Russia}
\author{F.~Batsch}
\affiliation{Max Planck Institute for Physics, Munich, Germany}
\author{M.~Bergamaschi}
\affiliation{Max Planck Institute for Physics, Munich, Germany}
\author{P.~Blanchard}
\affiliation{Ecole Polytechnique Federale de Lausanne (EPFL), Swiss Plasma Center (SPC), Lausanne, Switzerland}
\author{P.N.~Burrows}
\affiliation{John Adams Institute, Oxford University, Oxford, UK}
\author{B.~Buttensch{\"o}n}
\affiliation{Max Planck Institute for Plasma Physics, Greifswald, Germany}
\author{J.~Chappell}
\affiliation{UCL, London, UK}
\author{E.~Chevallay}
\affiliation{CERN, Geneva, Switzerland}
\author{M.~Chung}
\affiliation{UNIST, Ulsan, Republic of Korea}
\author{D.A.~Cooke}
\affiliation{UCL, London, UK}
\author{H.~Damerau}
\affiliation{CERN, Geneva, Switzerland}
\author{C.~Davut}
\affiliation{Cockcroft Institute, Daresbury, UK} 
\affiliation{University of Manchester, Manchester, UK}
\author{G.~Demeter}
\affiliation{Wigner Research Center for Physics, Budapest, Hungary}
\author{A.~Dexter}
\affiliation{Cockcroft Institute, Daresbury, UK} 
\affiliation{Lancaster University, Lancaster, UK}
\author{S.~Doebert}
\affiliation{CERN, Geneva, Switzerland}
\author{J.~Farmer}
\affiliation{CERN, Geneva, Switzerland}
\affiliation{Max Planck Institute for Physics, Munich, Germany}
\author{A.~Fasoli}
\affiliation{Ecole Polytechnique Federale de Lausanne (EPFL), Swiss Plasma Center (SPC), Lausanne, Switzerland}
\author{V.N.~Fedosseev}
\affiliation{CERN, Geneva, Switzerland}
\author{R.~Fiorito}
\affiliation{Cockcroft Institute, Daresbury, UK} 
\affiliation{University of Liverpool, Liverpool, UK}
\author{R.A.~Fonseca}
\affiliation{ISCTE - Instituto Universit\'{e}ario de Lisboa, Portugal} 
\affiliation{GoLP/Instituto de Plasmas e Fus\~{a}o Nuclear, Instituto Superior T\'{e}cnico, Universidade de Lisboa, Lisbon, Portugal}
\author{I.~Furno}
\affiliation{Ecole Polytechnique Federale de Lausanne (EPFL), Swiss Plasma Center (SPC), Lausanne, Switzerland}
\author{S.~Gessner}
\affiliation{CERN, Geneva, Switzerland}
\affiliation{SLAC, Menlo Park, CA, USA}
\author{A.A.~Gorn}
\affiliation{Budker Institute of Nuclear Physics SB RAS, Novosibirsk, Russia}
\affiliation{Novosibirsk State University, Novosibirsk, Russia}
\author{E.~Granados}
\affiliation{CERN, Geneva, Switzerland}
\author{M.~Granetzny}
\affiliation{University of Wisconsin, Madison, Wisconsin, USA}
\author{T.~Graubner}
\affiliation{Philipps-Universit{\"a}t Marburg, Marburg, Germany}
\author{O.~Grulke}
\affiliation{Max Planck Institute for Plasma Physics, Greifswald, Germany}
\affiliation{Technical University of Denmark, Lyngby, Denmark}
\author{E.~Gschwendtner}
\affiliation{CERN, Geneva, Switzerland}
\author{E.D.~Guran}
\affiliation{CERN, Geneva, Switzerland}
\author{J.R.~Henderson}
\affiliation{Cockcroft Institute, Daresbury, UK}
\affiliation{Accelerator Science and Technology Centre, ASTeC, STFC Daresbury Laboratory, Warrington, UK}
\author{M.~H{\"u}ther}
\affiliation{Max Planck Institute for Physics, Munich, Germany}
\author{M.{\'A}.~Kedves}
\affiliation{Wigner Research Center for Physics, Budapest, Hungary}
\author{V.~Khudyakov}
\affiliation{Heinrich-Heine-Universit{\"a}t D{\"u}sseldorf, D{\"u}sseldorf, Germany}
\affiliation{Budker Institute of Nuclear Physics SB RAS, Novosibirsk, Russia}
\author{S.-Y.~Kim}
\affiliation{UNIST, Ulsan, Republic of Korea}
\affiliation{CERN, Geneva, Switzerland}
\author{F.~Kraus}
\affiliation{Philipps-Universit{\"a}t Marburg, Marburg, Germany}
\author{M.~Krupa}
\affiliation{CERN, Geneva, Switzerland}
\author{T.~Lefevre}
\affiliation{CERN, Geneva, Switzerland}
\author{L.~Liang}
\affiliation{Cockcroft Institute, Daresbury, UK}
\affiliation{University of Manchester, Manchester, UK}
\author{N.~Lopes}
\affiliation{GoLP/Instituto de Plasmas e Fus\~{a}o Nuclear, Instituto Superior T\'{e}cnico, Universidade de Lisboa, Lisbon, Portugal}
\author{K.V.~Lotov}
\affiliation{Budker Institute of Nuclear Physics SB RAS, Novosibirsk, Russia}
\affiliation{Novosibirsk State University, Novosibirsk, Russia}
\author{S.~Mazzoni}
\affiliation{CERN, Geneva, Switzerland}
\author{D.~Medina~Godoy} 
\affiliation{CERN, Geneva, Switzerland}
\author{J.T.~Moody}
\affiliation{Max Planck Institute for Physics, Munich, Germany}
\author{K.~Moon}
\affiliation{UNIST, Ulsan, Republic of Korea}
\author{P.I.~Morales~Guzm\'{a}n}
\affiliation{Max Planck Institute for Physics, Munich, Germany}
\author{M.~Moreira}
\affiliation{GoLP/Instituto de Plasmas e Fus\~{a}o Nuclear, Instituto Superior T\'{e}cnico, Universidade de Lisboa, Lisbon, Portugal}
\author{T.~Nechaeva}
\affiliation{Max Planck Institute for Physics, Munich, Germany}
\author{E.~Nowak}
\affiliation{CERN, Geneva, Switzerland}
\author{C.~Pakuza}
\affiliation{John Adams Institute, Oxford University, Oxford, UK}
\author{H.~Panuganti}
\affiliation{CERN, Geneva, Switzerland}
\author{A.~Pardons}
\affiliation{CERN, Geneva, Switzerland}
\author{A.~Perera}
\affiliation{Cockcroft Institute, Daresbury, UK}
\affiliation{University of Liverpool, Liverpool, UK}
\author{J.~Pucek}
\affiliation{Max Planck Institute for Physics, Munich, Germany}
\author{A.~Pukhov}
\affiliation{Heinrich-Heine-Universit{\"a}t D{\"u}sseldorf, D{\"u}sseldorf, Germany}
\author{B.~R\'{a}czkevi}
\affiliation{Wigner Research Center for Physics, Budapest, Hungary}
\author{R.L.~Ramjiawan}
\affiliation{CERN, Geneva, Switzerland}
\affiliation{John Adams Institute, Oxford University, Oxford, UK}
\author{S.~Rey}
\affiliation{CERN, Geneva, Switzerland}
\author{O.~Schmitz}
\affiliation{University of Wisconsin, Madison, Wisconsin, USA}
\author{E.~Senes}
\affiliation{CERN, Geneva, Switzerland}
\author{L.O.~Silva}
\affiliation{GoLP/Instituto de Plasmas e Fus\~{a}o Nuclear, Instituto Superior T\'{e}cnico, Universidade de Lisboa, Lisbon, Portugal}
\author{C.~Stollberg}
\affiliation{Ecole Polytechnique Federale de Lausanne (EPFL), Swiss Plasma Center (SPC), Lausanne, Switzerland}
\author{A.~Sublet}
\affiliation{CERN, Geneva, Switzerland}
\author{A.~Topaloudis}
\affiliation{CERN, Geneva, Switzerland}
\author{N.~Torrado}
\affiliation{GoLP/Instituto de Plasmas e Fus\~{a}o Nuclear, Instituto Superior T\'{e}cnico, Universidade de Lisboa, Lisbon, Portugal}
\author{P.V.~Tuev}
\affiliation{Budker Institute of Nuclear Physics SB RAS, Novosibirsk, Russia}
\affiliation{Novosibirsk State University, Novosibirsk, Russia}
\author{M.~Turner}
\affiliation{CERN, Geneva, Switzerland}
\affiliation{LBNL, Berkeley, CA, USA}
\author{F.~Velotti}
\affiliation{CERN, Geneva, Switzerland}
\author{L.~Verra}
\affiliation{Max Planck Institute for Physics, Munich, Germany}
\affiliation{CERN, Geneva, Switzerland}
\affiliation{Technical University Munich, Munich, Germany}
\author{J.~Vieira}
\affiliation{GoLP/Instituto de Plasmas e Fus\~{a}o Nuclear, Instituto Superior T\'{e}cnico, Universidade de Lisboa, Lisbon, Portugal}
\author{H.~Vincke}
\affiliation{CERN, Geneva, Switzerland}
\author{C.P.~Welsch}
\affiliation{Cockcroft Institute, Daresbury, UK}
\affiliation{University of Liverpool, Liverpool, UK}
\author{M.~Wendt}
\affiliation{CERN, Geneva, Switzerland}
\author{M.~Wing}
\affiliation{UCL, London, UK}
\author{J.~Wolfenden}
\affiliation{Cockcroft Institute, Daresbury, UK}
\affiliation{University of Liverpool, Liverpool, UK}
\author{B.~Woolley}
\affiliation{CERN, Geneva, Switzerland}
\author{G.~Xia}
\affiliation{Cockcroft Institute, Daresbury, UK}
\affiliation{University of Manchester, Manchester, UK}
\author{M.~Zepp}
\affiliation{University of Wisconsin, Madison, Wisconsin, USA}
\author{G.~Zevi~Della~Porta}
\affiliation{CERN, Geneva, Switzerland}
\collaboration{The AWAKE Collaboration}

\noaffiliation

\sloppy

\begin{abstract}
A precise characterization of the incoming proton bunch parameters is required to accurately simulate the self-modulation process in the Advanced Wakefield Experiment (AWAKE). This paper presents an analysis of the parameters of the incoming proton bunches used in the later stages of the AWAKE Run 1 data-taking period. The transverse structure of the bunch is observed at multiple positions along the beamline using scintillating or optical transition radiation screens. The parameters of a model that describes the bunch transverse dimensions and divergence are fitted to represent the observed data using Bayesian inference. The analysis is tested on simulated data and then applied to the experimental data.

\end{abstract}

\maketitle


\section{Introduction}

Plasma can sustain high electric fields and can be used to produce accelerating gradients larger than in conventional particle accelerators~\cite{tajima1979laser, chen1985acceleration, blumenfeld2007energy, gonsalves2019petawatt, albert20212020}.   
The Advanced Wakefield Experiment (AWAKE) \cite{caldwell2013awake} is an experiment located at CERN that investigates proton-driven plasma wakefield acceleration~\cite{caldwell2009proton}. AWAKE uses proton bunches from the CERN Super Proton Synchrotron (SPS) with the energy of \SI{400}{\giga\electronvolt} to accelerate a witness bunch of electrons in a ten-meter-long rubidium plasma source. 
The seeded self-modulation mechanism~\cite{kumar2010self, muggli2017awake} is used to divide the long driver bunch into a group of shorter bunches that can resonantly drive wakefields~\cite{adli2019experimental, batsch2021transition}. 

\begin{figure*}[!t]
\includegraphics[width=0.8\textwidth]{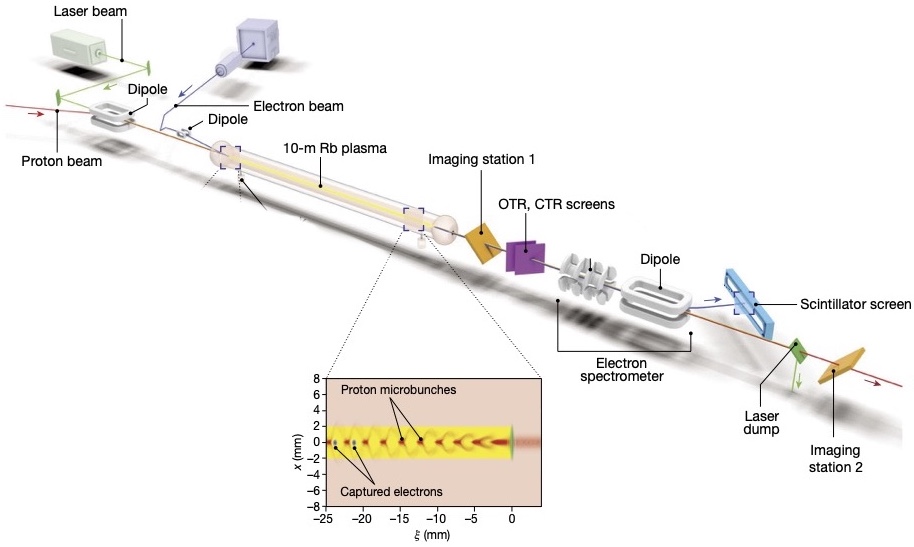}
\caption{The layout of the AWAKE experiment. Electron, proton, and laser beams that propagate from left to right are shown in blue, red, and green colors, respectively. The bottom subplot shows the self-modulated proton bunch at the plasma exit. Retrieved from \cite{adli2018acceleration}.}
\label{fig:layout}
\end{figure*}

A detailed understanding of the parameters of the incoming proton bunches is important for the understanding of the experimental data. The bunch and plasma parameters define the wakefield structure, and a comparison of experimental observations to simulations requires correct bunch parameters.  It has been shown \cite{gorn2020proton} that results of plasma modeling are sensitive to the parameters of the driver bunch, and uncertainty in the input bunch parameters complicates the comparison of the simulation results and experimental measurements.

In this paper, we present a determination of the parameters of the unmodulated proton bunch by combining the data from multiple beam imaging systems. We test our analysis using simulated data and then apply it to experimental data to study variations of the parameters over a large number of events.

The paper is structured in the following way. In section~\ref{sec:2}, a short overview of the AWAKE experiment is given with a focus on the proton bunch, as it plays the central role in our study. In section~\ref{sec:3}, the experimental setup used in our analysis is described, an overview of the acquired data is given and preliminary investigations are discussed. A statistical model is presented and tested on simulated data in section~\ref{sec:4}. This is followed by section~\ref{sec:5}, in which parameters and their prior probabilities are discussed, and results are presented. Finally, conclusions are presented in section~\ref{sec:6}.  

\section{Proton Bunch in the AWAKE experiment}
\label{sec:2}

The proton bunches used in the AWAKE experiment are produced in the CERN accelerator complex. They are accelerated in the Linac 4, Proton Synchrotron Booster, and Proton Synchrotron accelerating stages and reach the required energy of \SI{400}{\giga\electronvolt} in the SPS. After this, they are sent to the AWAKE facility with intervals of $15-30~\si{\second}$. Extraction of a proton bunch to the AWAKE facility will be further denoted as an ‘event’. 

In AWAKE, the proton bunch enters a ten-meter-long rubidium vapor section~\cite{oz2014novel, plyushchev2017rubidium} together with the co-propagating laser pulse (see Fig.~\ref{fig:layout}). According to the Run 1 baseline~\cite{gschwendtner2016awake}, the bunch is focused at $ z_w = (5 \pm 3) \; \si{\centi\meter}$ after the entrance to the rubidium section. The radial size is  $\sigma_{x,y} = (0.20 \pm 0.04) \;  \si{\milli\meter}$, longitudinal size is $ \sigma_z = 6-8 \; \si{\centi\meter}$ and angular divergence is $\sigma'_{x,y} = (4 \pm 2) \times \SI{d-5}{\radian}$, where $\sigma$ represents a Gaussian standard deviation. The parameters of proton bunches, such as the bunch centroid and population, fluctuate from event to event.  

The laser pulse~\cite{fedosseev2016integration} with a maximum energy of \SI{450}{\milli\joule} and a central wavelength of \SI{780}{\nano\meter} ionizes the rubidium vapor, creating a relativistic ionization front that co-propagates with the proton bunch. The relativistic ionization front is much shorter than the typical period of the wakefields ($>3~\si{\pico\second}$) and it is used to seed the controlled self-modulation instability of the proton bunch. The resulting microbunches then act resonantly to drive large wakefields during the plasma section.

In the SPS accelerating ring, wire scanners are used to measure the transverse and longitudinal bunch emittances with the accuracy of 20\%~\cite{caldwell2013awake, berrig2014cern, baud2013performance}. The Beam Quality Monitors~\cite{papotti2008beam, papotti2011sps} are used to measure longitudinal bunch profiles, and Beam Current Transformers (BCT)~\cite{monera2011upgrade} are used to determine the bunch intensity during the whole accelerating cycle. 

In the AWAKE experiment, transverse and longitudinal structures of the bunch can be observed before and after the plasma section using diagnostics based on Optical Transition Radiation (OTR) or scintillating light, which is produced when the bunch crosses light-emitting screens placed at $45^{\circ}$ to the beamline ~\cite{caldwell2013awake}.

\section{Measurements and Data}
\label{sec:3}

\begin{figure}[!t]
\includegraphics{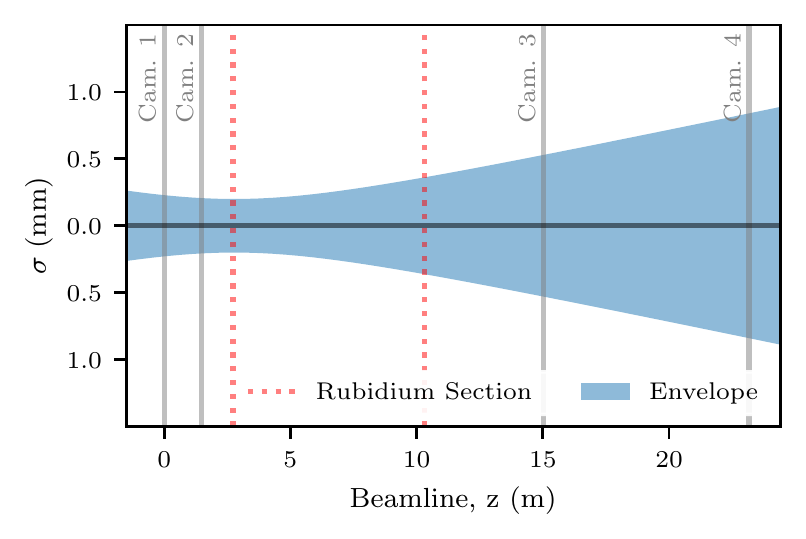}
\caption{The standard deviation (envelope) of the transverse proton bunch profile versus the beamline position for nominal bunch parameters without plasma. Gray solid lines show positions of four beam observation systems. The position of the plasma section is shown by red dotted lines.}
\label{fig:setup-example}
\end{figure}

\begin{figure*}[!t]
\includegraphics{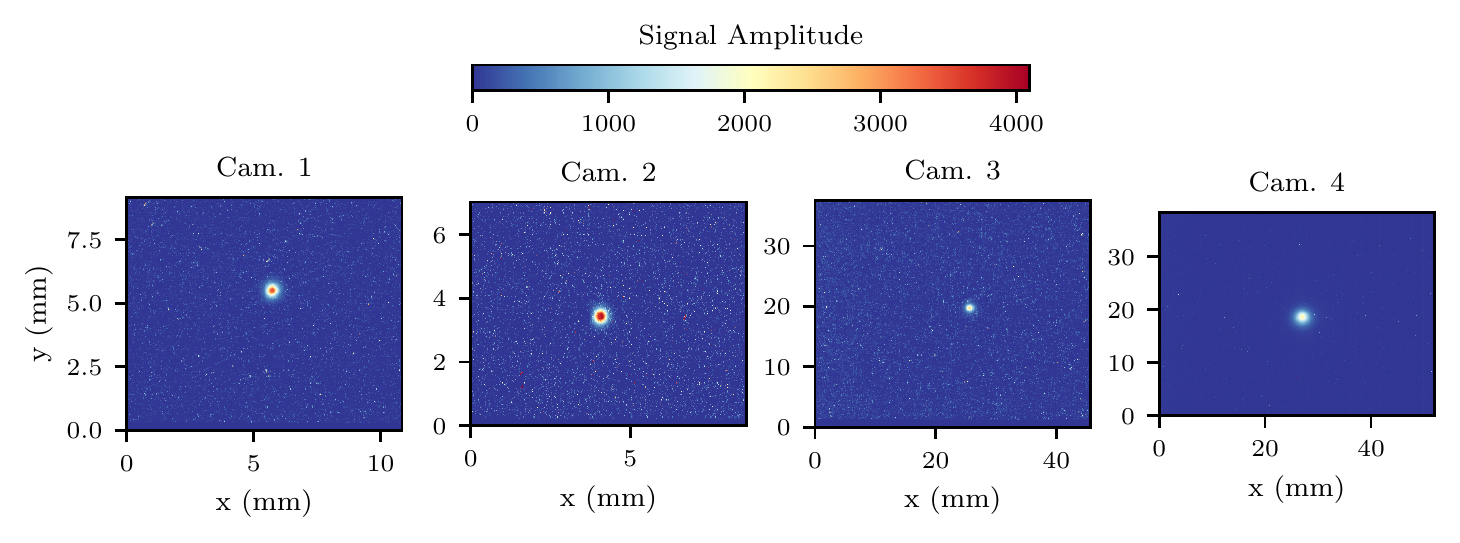}
\caption{Images from one event with the proton bunch population $n=24.04\times10^{10}p^+$. Subplots correspond to the four beam imaging systems specified in Table~\ref{table:setup-cam}. Signal amplitude represents the value recorded by the camera pixel, and it is in the range $[0, 4095]$. }
\label{fig:event-example}
\end{figure*}

\begin{table}[!b]
\caption{Description of the beam imaging systems (denoted as Cam.~1-4) used in the measurements. The position along the beamline is denoted as $z$, and it is measured from the position of the first camera. }
\begin{ruledtabular}
\begin{tabular}{lccccc}
 & \textbf{Cam. 1} & \textbf{Cam. 2}  & \textbf{Cam. 3} & \textbf{Cam. 4} \\
 \hline
 \textbf{Position}, $z$ (m)  &  0.000  & 1.478    &     15.026        &  23.164        \\ 
\textbf{Screen Type} &  OTR  & OTR   &   OTR       &    Scint.         \\ 
\textbf{Screen Material} &  Si Ag  & Si Ag   &   Si Ag    &    Chromox         \\ 
\textbf{Camera Type} &  CCD  & CCD   &   CCD    &    CMOS         \\ 
\textbf{\# Pixels} &  $400\times300$  & $400\times300$   &   $400\times300$    &    $1280\times960$  \\ 
\end{tabular}
\end{ruledtabular}
\label{table:setup-cam}
\end{table}

Four beam imaging systems that capture images of the transverse profile of the unmodulated proton bunch before and after the rubidium vapor section were used in our analysis. Parameters of the beam imaging systems are summarized in Table~\ref{table:setup-cam}. The first three stations have CCD cameras with OTR screens, and the last station has a digital CMOS camera with a scintillating screen. Fig.~\ref{fig:setup-example} illustrates the relative positions of the measurement stations, the plasma section, and the envelope trajectory of the unmodulated bunch for the baseline parameters. It can be seen that the first two stations are located very close to the waist position, so they mainly carry information about the size of the bunch close to the focus. The last two stations are located much farther from the waist position, and they are primarily sensitive to the angular divergence of the bunch.  

We performed measurements on October 10, 2018, during which a dataset that consists of $672$ events was collected.  By ‘event data’ we denote 4 images from the beam imaging systems and the proton bunch population measured in the SPS using the BCT. Four types of proton bunches were requested from the SPS operators with the parameters summarized in Table~\ref{table:datasets}. The first two types correspond to the bunches with small, $(7.77-10.30)\times10^{10} p^+$, and large, $(23.20 - 28.00)\times10^{10} p^+$,  populations. These events are intended to study the impact of the bunch population on the bunch emittance and focal size. The second two types represent bunches with or without longitudinal compression. The bunch compression is achieved via a rotation in longitudinal phase space using a voltage step with a fast rise time~\cite{timko2013short, bracco2014beam}. 
The typical longitudinal bunch length (rms) without the bunch rotation is $\approx\SI{9.6}{\centi\meter}$ and with the bunch rotation $\approx\SI{7.9}{\centi\meter}$.
The acquired dataset represents proton bunches commonly used in the later stages of Run 1 data-taking period.

An example of the event data with a population of $n=24.04\times10^{10}p^+$ is shown in Fig.~\ref{fig:event-example}. The central part of the images represents the OTR light on the first three screens and the scintillating light on the last screen produced by the proton bunch. The background noise is produced primarily by secondary particles that are generated upstream of AWAKE. 

\begin{table}[!b]
\caption{Summary of the datasets. Events are divided into 4 categories, i.e., small and large bunch population and bunch rotation ON and OFF.}
\begin{ruledtabular}
\begin{tabular}{lcccc}
 \textbf{Symbol} & \textbf{$n$} $[10^{10}, p^+]$  & \textbf{Rotation} & \textbf{\# Events}\\ 
 \hline
$D_{11}$  &   7.77 - 10.30  & ON     &     181    \\ 
$D_{12}$  & 7.77 - 10.30    &  OFF   &     160    \\ 
$D_{21}$  & 23.20 - 28.00   &  ON    &     139    \\ 
$D_{22}$. & 23.20 - 28.00   &  OFF   &     192    \\ 
\end{tabular}
\end{ruledtabular}
\label{table:datasets}
\end{table}

\subsection{Preliminary Investigations}

The analysis relies on knowledge of calibration factors and resolutions scales for the devices used.  These are described briefly.

\subsubsection{Pixel Calibration Factor}

To determine the calibration factor of the pixel sizes, we use calibration frames that are engraved on the surface of each light-emitting screen.  Parameters of the frames, such as width, height, and sizes of the engraving lines, are known to high precision. The screens are placed at an angle of \SI{45}{\degree} with respect to the beamline and rotated on the horizontal axis. To calculate the horizontal calibration factor, we divide the absolute size of the frame by the number of pixels that correspond to it. To determine the vertical calibration factor, the absolute size of the frame is multiplied by $\cos(\pi/4)$ and then divided by the number of corresponding pixels. In the following, we assume that the pixel size includes the correct calibration factors. The resulting pixel sizes for each camera are specified in Table~\ref{tab:posterior-summary}. 

\subsubsection{Resolution Function}

There are two types of resolution effects present in the measurements. The first is the resolution of the optical system of the camera. The second is the resolution of the light-emitting screen. We assume that they both are convolutions with Gaussian kernels and their superposing effect is also represented by Gaussian convolution with the variance given by the quadrature sum of these two components. 

We estimate the optical camera resolution using the images of the calibration frame previously discussed in the pixel calibration. For this, we assume that the engravings have sharp edges and can be represented as a product of Heaviside step functions. The images of the frames captured by the corresponding cameras are used to determine the unknown optical resolution parameters.

The study~\cite{burger2016scintillation} performed at the CERN HiRadMat test facility indicates that scintillating screens are characterized by a worse resolution compared to the OTR screens.  We include the additional smearing from the scintillating screen in our analysis using nuisance variables and consider constant resolution parameters for 3 cameras with OTR screens. The detailed values of resolution parameters are specified in Table~\ref{tab:posterior-summary}.

\subsubsection{Background Distribution}

\begin{figure}[b]
\includegraphics{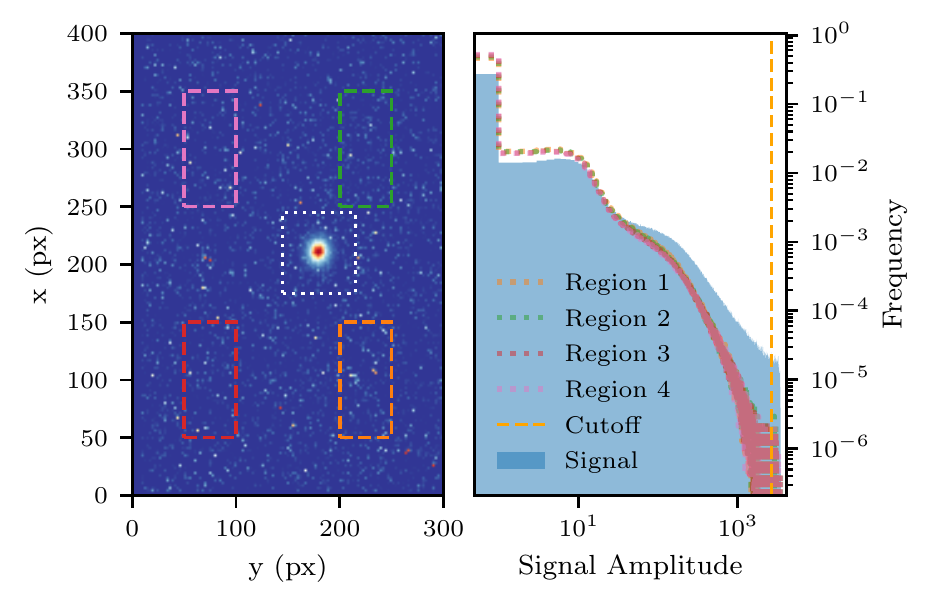}
\caption{Background distributions from different regions of Cam. 1. The frequency represents the number of counts present in each bin normalized such that the sum of frequencies from each bin is equal to one. The pixels enclosed by the white dotted lines are used to determine the parameters of the proton bunch. The background distributions from four rectangular regions enclosed by the dashed lines are shown in the right subplot. The dashed line in the right subplot shows the truncation threshold. The filled region represent the distribution of the signal (enclosed by the white dotted line). }
\label{fig:background-data}
\end{figure}

\begin{figure}[b]
\includegraphics{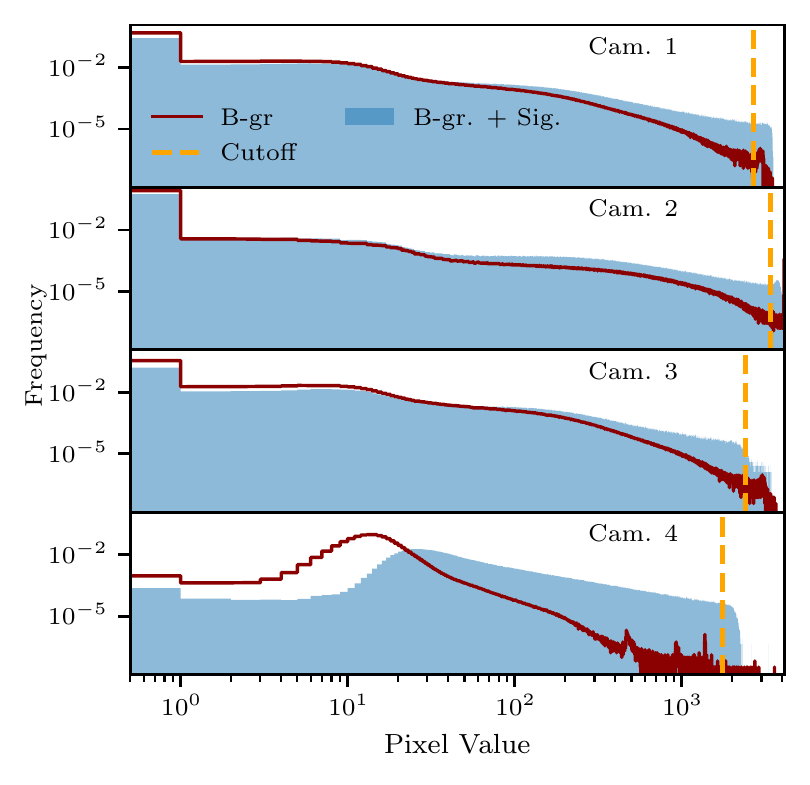}
\caption{Background distributions corresponding to the dataset with a large bunch population. The dark red colors show the distribution of the background only. The filled histograms show the distributions from pixels used in the analysis of bunch parameters, and they represent the superposition of signal and background. Orange lines show positions of the cutoff values. }
\label{fig:background-cams}
\end{figure}

We divide the image from each camera into different regions (see Fig.~\ref{fig:background-data}). Those pixels located in the central region --- which is approximately 4-5 standard deviations around the bunch centroid --- are used to infer the parameters of the proton bunch. The remaining pixels are combined from multiple events to approximate the background distribution via binned histograms. We use separate background distributions for each camera and datasets with small and large bunch populations. An example of the background distributions from four different regions of one camera is shown in Fig.~\ref{fig:background-data}. It can be seen that all histograms significantly overlap, demonstrating that the background follows a similar distribution in different parts of the camera. 

Background distributions from the four different cameras are shown in Fig.~\ref{fig:background-cams}. The histograms corresponding to cameras with the OTR screens have a similar structure. Namely, they all have a maximum at 0, long tails that extend to the amplitudes of $>2000$, and a small bump of saturated pixels at the righthand side of the histograms. To avoid a possible negative impact of the saturation on the analysis, we discard those pixels that exceed the threshold values of $2700$, $3400$, $2400$,$1750$, for each camera, respectively (see also dashed lines in Fig.~\ref{fig:background-cams}). The histograms of signal and background show a smooth behavior up to these threshold values without signs of saturation. 

The images produced by the last beam observing system have  $\sim10$ times more pixels compared to the images from the three other cameras. To improve the run-time of the analysis, we average every $3\times3$ pixels from the last camera.  Averaging of the pixels reduces the long tail of the distribution as can be seen in Fig.~\ref{fig:background-cams} (bottom subplot). In the following, we will assume that the background on each pixel of the last camera is well-modeled with a truncated (form $0$ to $4095$) Gaussian distribution. The distribution variance is determined from the histogram, and the mean is fitted with a free parameter.

\section{Statistical Model}
\label{sec:4}

We perform the statistical analysis using a Bayesian approach. Prior probabilities about parameters $\boldsymbol{\lambda}, \boldsymbol{\nu}$ of the model $M$ are updated to posterior probabilities using Bayes’ theorem 
\begin{equation}
\label{eq:bayes}
P(\boldsymbol{\lambda}, \boldsymbol{\nu}|M, D) = \frac{P(D| \boldsymbol{\lambda}, \boldsymbol{\nu}) \cdot P(\boldsymbol{\lambda}, \boldsymbol{\nu}|M)}{\int P(D| \boldsymbol{\lambda}, \boldsymbol{\nu}) \cdot P(\boldsymbol{\lambda}, \boldsymbol{\nu}|M) d\boldsymbol{\lambda} d\boldsymbol{\nu}},
\end{equation}
where $P(D| \boldsymbol{\lambda}, \boldsymbol{\nu}) $ is a likelihood that represents a probability of data given the model, $P(\boldsymbol{\lambda}, \boldsymbol{\nu}|M)$ is a prior, and $P(\boldsymbol{\lambda}, \boldsymbol{\nu}|M, D)$ is a posterior probability distribution. The data from one event is denoted as $D \equiv \left \{ d_{x,y}^j  \right \}$, where $d$ is a signal from one pixel, $x,y$ are the row, column of the pixel and $j$ represents the camera index. The dataset consists of multiple events $\left \{ D \right \}_i$ where $i$ denotes the event index. In Eq.~\ref{eq:bayes}, $\boldsymbol{\lambda}$ represents parameters of interest, and $\boldsymbol{\nu}$ represents nuisance parameters. Some of the nuisance parameters are kept constant, and others are free parameters of the fit that will be marginalized at the final stage of the analysis.  

We analyze events consecutively and the likelihood of one event is defined as a product of likelihoods of individual pixels
\begin{equation}
\label{eq:ll2}
P(D| \boldsymbol{\lambda}, \boldsymbol{\nu}) = \prod_{j \in N_{cam}} \prod_{y \in N_{rows}} \prod_{x \in N_{columns}}  p(d_{xy}^j|\boldsymbol{\lambda}, \boldsymbol{\nu}),
\end{equation}
where $p(d_{xy}^j|\boldsymbol{\lambda}, \boldsymbol{\nu})$ is the probability of the detected signal for one pixel given the model parameters. 

\begin{figure}[!t] 
    \centering
    \includegraphics{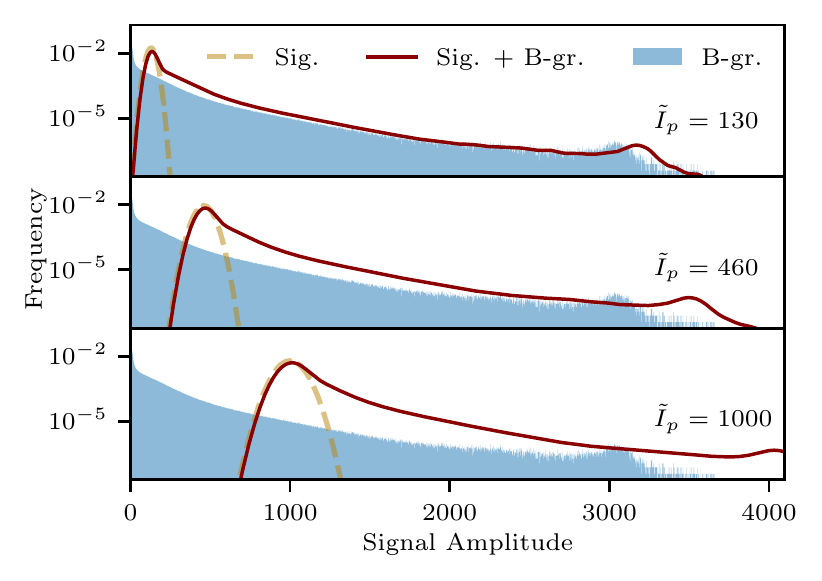}
    \caption{Convolutions of the probability distribution of the signal created by the proton bunch and the background distribution are shown for different assumed mean signal amplitudes and variances for Cam.~3. The signal from the proton bunch, $P_p(d|\boldsymbol{\lambda}, \boldsymbol{\nu})$, is assumed to be normally distributed (shown in dashed lines). The filled histograms show the background noise. The red lines show a numerical convolution represented by $p(d|\boldsymbol{\lambda}, \boldsymbol{\nu})$. }
    \label{fig:signal-back-conv-2}
\end{figure}

The $d_{xy}^j$ are composed of a background noise, with a probability distribution denoted as $P_b(d_{xy}^j)$, and a signal $P_p(d_{xy}^j|\boldsymbol{\lambda}, \boldsymbol{\nu})$ produced by the proton bunch. In the following, we will avoid indices $x, y, j$ in $d_{xy}^j$ for notational convenience if one pixel is considered.
We assume that, for the given camera, the background is the same for all pixels and we approximate it by a binned histogram (see previous section). The resulting superposition of these two contributions is given by the convolution 
\begin{equation}
\label{eq:ll3}
p(d|\boldsymbol{\lambda}, \boldsymbol{\nu}) = \int P_p(\tau|\boldsymbol{\lambda}, \boldsymbol{\nu}) \cdot P_b(d - \tau) d\tau.
\end{equation}
The convolution is computed numerically for the first three cameras due to the non-analytic form of the background, and it is computed analytically for the last camera, where the background is Gaussian. An example for such a convolution is shown in Fig.~\ref{fig:signal-back-conv-2}, where it is assumed that the signal from the proton bunch has a Gaussian distribution.  The evaluation of the numerical convolutions is computationally expensive, but it is needed to include correctly the non-analytic features of the background.

\subsection{Bunch Propagation Equation}

We consider a model in which each particle of the bunch follows a linear equation of motion defined as 
\begin{equation}
\label{eq:env-eq}
\mathbf{r}_i = \mathbf{r_0}_i +  \boldsymbol{r}'_i \left ( z - z_{w} \right ),
\end{equation} 
where $i$ denotes the particle's index, $z_w$ denotes the waist position, i.e., the coordinate along the beamline in which the radial bunch size is minimal, $\mathbf{r}_{0} = (x_{0}, y_{0})$ is the particle position at the waist, and $\boldsymbol{r}' = (dx/dz, dy/dz) \approx (\theta_{xz}, \theta_{yz})$ is the particle angle with respect to the beamline in the $x-z$ and $y-z$ planes. The distance along the beamline is denoted by $z$ and is defined as the trajectory of the bunch centroid. 

We measure distances in the transverse directions relative to the center of the proton bunch. To determine the center of the bunch, we use two parameters per camera, one for the $x$ and one for the $y$ directions, labeled as $\mu_{j, x}, \mu_{j, y}$ with $j$ denoting the camera index (see Table~\ref{tab:posterior-summary}). 

The envelope equation that describes the transverse size of such a bunch along the beamline can be defined as
\begin{equation}
\label{eq:env-eq2}
\begin{split}
\boldsymbol{\sigma}^2 & = \left \langle \boldsymbol{r}^2 \right \rangle \\
& = \left \langle \boldsymbol{r}_0^2 \right \rangle + 2 \left \langle \boldsymbol{r_0}\boldsymbol{r}' \right \rangle (z-z_w) + \left \langle \boldsymbol{r}'^2 \right \rangle (z-z_w)^2,
\end{split}
\end{equation} 
where $\left \langle \cdot \right \rangle$ denotes the average over the ensemble of particles. It is assumed that there is no correlation between the $x$ and $y$ projections of the bunch, and that $\boldsymbol{r_0} $ and $\boldsymbol{r}'$ are not correlated at the waist position, i.e., $ \left \langle \boldsymbol{r_0}\boldsymbol{r}' \right \rangle = 0$. For notational convenience, the bunch size at the waist will be denoted as $\boldsymbol{\sigma}_0 = \sqrt{\left \langle \boldsymbol{r}_0^2 \right \rangle}$, and the angular spread of the bunch as $\boldsymbol{\sigma'} = \sqrt{\left \langle \boldsymbol{r}'^2 \right \rangle} $. 

We consider two models that describe the distribution of protons in the beamline. 

In the first model, we assume that the proton bunch is represented by a single Gaussian distribution in the transverse and longitudinal directions. The transverse distribution of the protons at each light-emitting screen can be described by a bivariate Gaussian distribution with the bunch size, $\boldsymbol{\sigma}(\boldsymbol{\lambda}, \boldsymbol{\nu})$, described by Eq.~\ref{eq:env-eq2}
\begin{equation}
\label{eq:env-eq4}
I_p(x,y|\boldsymbol{\lambda}, \boldsymbol{\nu}, M_1) = \mathcal{N}(x, y|\boldsymbol{\sigma}(\boldsymbol{\lambda}, \boldsymbol{\nu}), \boldsymbol{\mu}(\boldsymbol{\lambda}, \boldsymbol{\nu})),
\end{equation}
and this model will be further denoted as ‘Single Gaussian’.

In the second model, denoted as ‘Double Gaussian’, we assume that the proton bunch is represented by a mixture of two Gaussian components. These components have the same alignment but different sizes, waist positions, and angular divergence. In this model, the transverse distribution of the protons at the light-emitting screen is defined as
\begin{eqnarray}
I_p(x,y|\boldsymbol{\lambda}, \boldsymbol{\nu}, M_2) = \alpha \mathcal{N}(x, y|\boldsymbol{\sigma_1}(\boldsymbol{\lambda}, \boldsymbol{\nu}), \boldsymbol{\mu_1}(\boldsymbol{\lambda}, \boldsymbol{\nu})) +\nonumber\\ 
+ (1-\alpha) \mathcal{N}(x, y|\boldsymbol{\sigma_2}(\boldsymbol{\lambda}, \boldsymbol{\nu}), \boldsymbol{\mu_1}(\boldsymbol{\lambda}, \boldsymbol{\nu}))
\label{eq:env-eq5}
\end{eqnarray}
where $\alpha$ controls the significance of each contribution. 
The component with larger angular divergence will be called ‘halo’ and the smaller ‘core’, and their parameters will be denoted by subscripts ‘c’ and ‘h’, respectively. The single and double Gaussian models are nested, and they predict the same result if $\alpha=1$. 

\subsection{Camera Modeling}

Light with intensity proportional to the number of particles in the bunch is emitted when the proton bunch crosses the light-emitting screen. This light experiences optical smearing that can be represented by a convolution of $I_p(x,y|\boldsymbol{\lambda}, \boldsymbol{\nu})$ with a Gaussian kernel $\mathcal{N}(x,y|\tilde{\sigma})$ with zero mean and resolution parameters $\tilde{\sigma}$, i.e.
\begin{eqnarray}
\label{eq:lf-1}
I_p(x,y|\boldsymbol{\lambda}, \boldsymbol{\nu})  = \int_{-\infty}^{\infty} \int_{-\infty}^{\infty} I_p(x - \tau_1,y- \tau_2|\boldsymbol{\lambda}, \boldsymbol{\nu})\times\nonumber\\ 
\times \mathcal{N}(\tau_1, \tau_2|\tilde{\sigma}) d \tau_1 d \tau_2.
\end{eqnarray}
The amount of light captured by one pixel is given by the integral over the pixel surface. We assume that the signal at each pixel is described by a Gaussian probability distribution with the mean given by $i_jI$, and the standard deviation of $f_j\sqrt{I}$, where $j$ denotes the camera index and $i_j$, $f_j$ are coefficients of proportionality represented by free parameters (see Table~\ref{tab:posterior-summary}):
\begin{equation}
\label{eq:lf-2}
P_p(d|\boldsymbol{\lambda}, \boldsymbol{\nu}) = \mathcal{N}(d | \mu = i_jI, \sigma = f_j\sqrt{I}).
\end{equation} 
This expression is used in Eq.~\ref{eq:ll3} to compute the likelihood for one pixel. The likelihoods from all pixels are multiplied together to get the final event likelihood described by Eq.~\ref{eq:ll2}. 

\begin{figure*}[!t]
\includegraphics{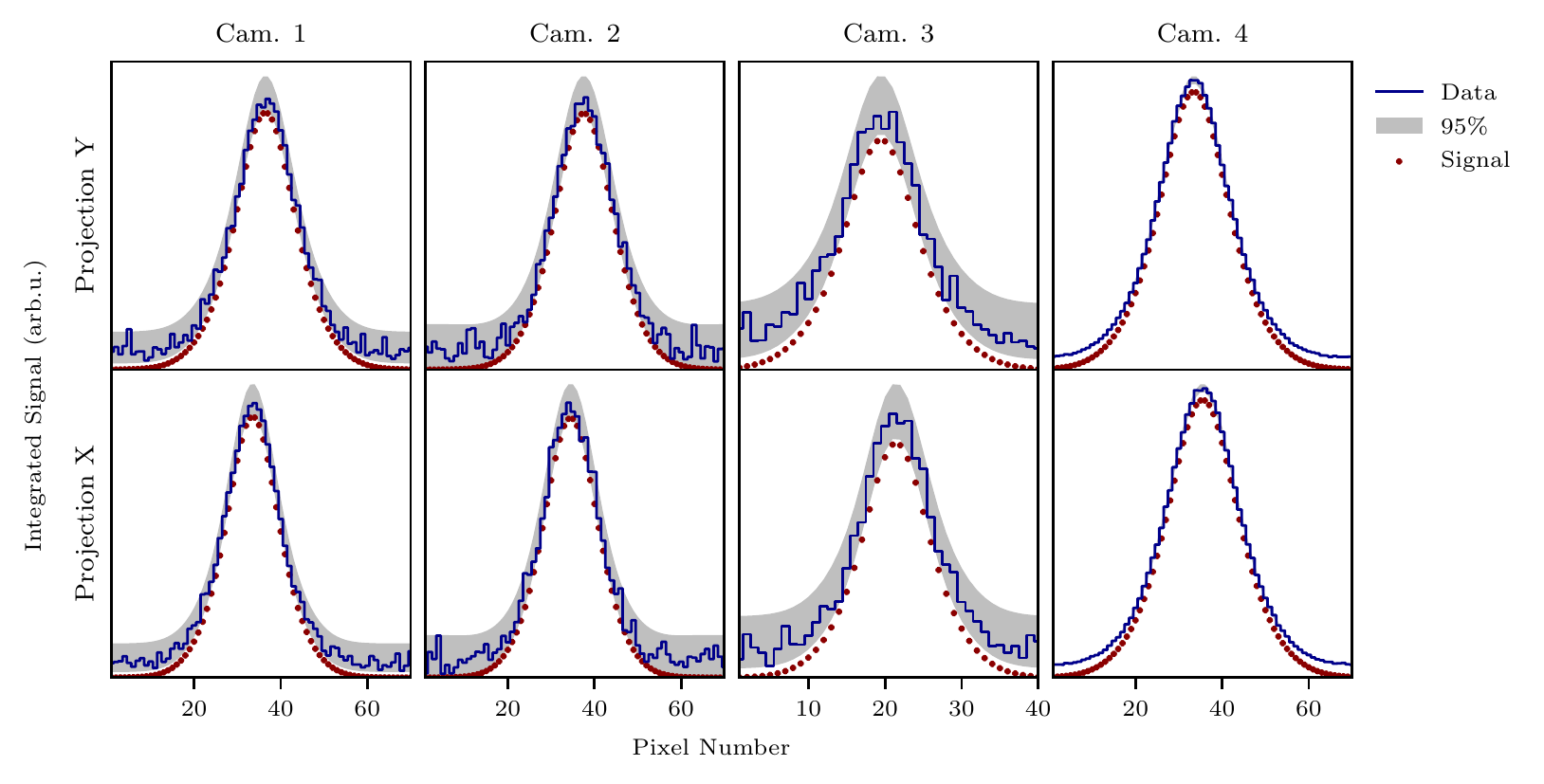}
\caption{A comparison of the simulated data and the best-fit result is shown for the double Gaussian model. The blue step-lines show an integrated signal over rows/columns of the camera. The grey-filled regions show the 95\% central probability intervals of the model including background and signal. The signal from the proton bunch is shown as points. The grey band for  Cam.~4 overlaps with the data and is not visible.}
\label{fig:toy-fit-2}
\end{figure*}

\begin{figure}[b]
\includegraphics{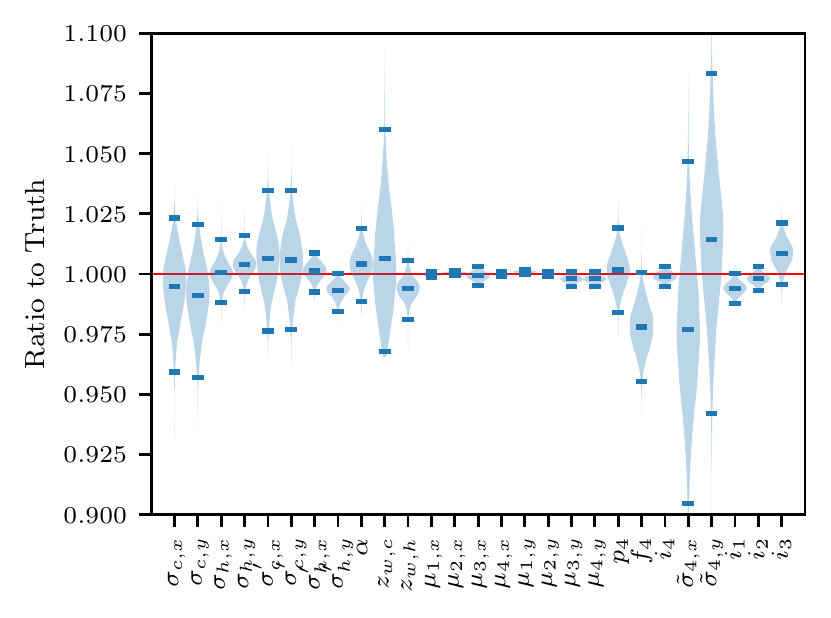}
\caption{The figure illustrates the posterior distribution obtained from the simulated event analysis. Each parameter is divided by truth to standardize the scale of the error bars. Blue horizontal ticks show 95\% central probability intervals and means. }
\label{fig:toy-params-2}
\end{figure}

We sample the posterior distribution given by Eq.~\ref{eq:bayes} using a Markov chain Monte Carlo (MCMC) algorithm. The likelihood is implemented in the Julia programming language, and the BAT.jl~\cite{schulz2021bat} package is used for the statistical inference. 

\subsection{Model Validation}

To validate that the analysis leads to the correct reconstruction of the parameters, we performed the following procedure:  
\begin{enumerate}
    \item True parameters of the models were defined.  
    \item Simulated events were generated that correspond to these parameters. The simulated events include background noise, lights fluctuations, optical smearing of cameras. 
    \item The analysis algorithms were applied and the resulting parameters were compared to the true values. 
\end{enumerate}

The validation procedure was performed for both single and double Gaussian models.  Excellent agreement was found between the extracted and true parameters for both models. 
An example of the simulated data and the fitted model for the double Gaussian bunch density is presented in Fig.~\ref{fig:toy-fit-2}. A violin plot with the parameter distributions is shown in Fig.~\ref{fig:toy-params-2}. It can be seen that true parameters are well within the 95\% central interval of the posterior distribution. 

Fig.~\ref{fig:toy-params-2} shows that some of the parameters, such as the alignment of the bunch, can be reconstructed with great accuracy, with uncertainties smaller than 3\% of pixel sizes. Another group of parameters, such as the transverse size and angular divergence, is characterized by uncertainties of a few percent. The nuisance parameters that describe the resolution function of the scintillating screen have the largest uncertainty. The camera with the scintillating screen is located at the largest distance from the waist position, and the bunch size is determined primarily by the angular divergence of the beam. The resolution is not a critical parameter, and it is not correlated significantly with the proton bunch parameters.  

We have tested the sensitivity of the analysis to the truncation of the data. Pixels that exceed the threshold values determined from the experimental data were discarded from the analysis, and the change in the resulting posterior means is well within the uncertainties of the parameter values. 

\section{Analysis}
\label{sec:5}

\begin{figure*}[!t]
\subfloat{\includegraphics{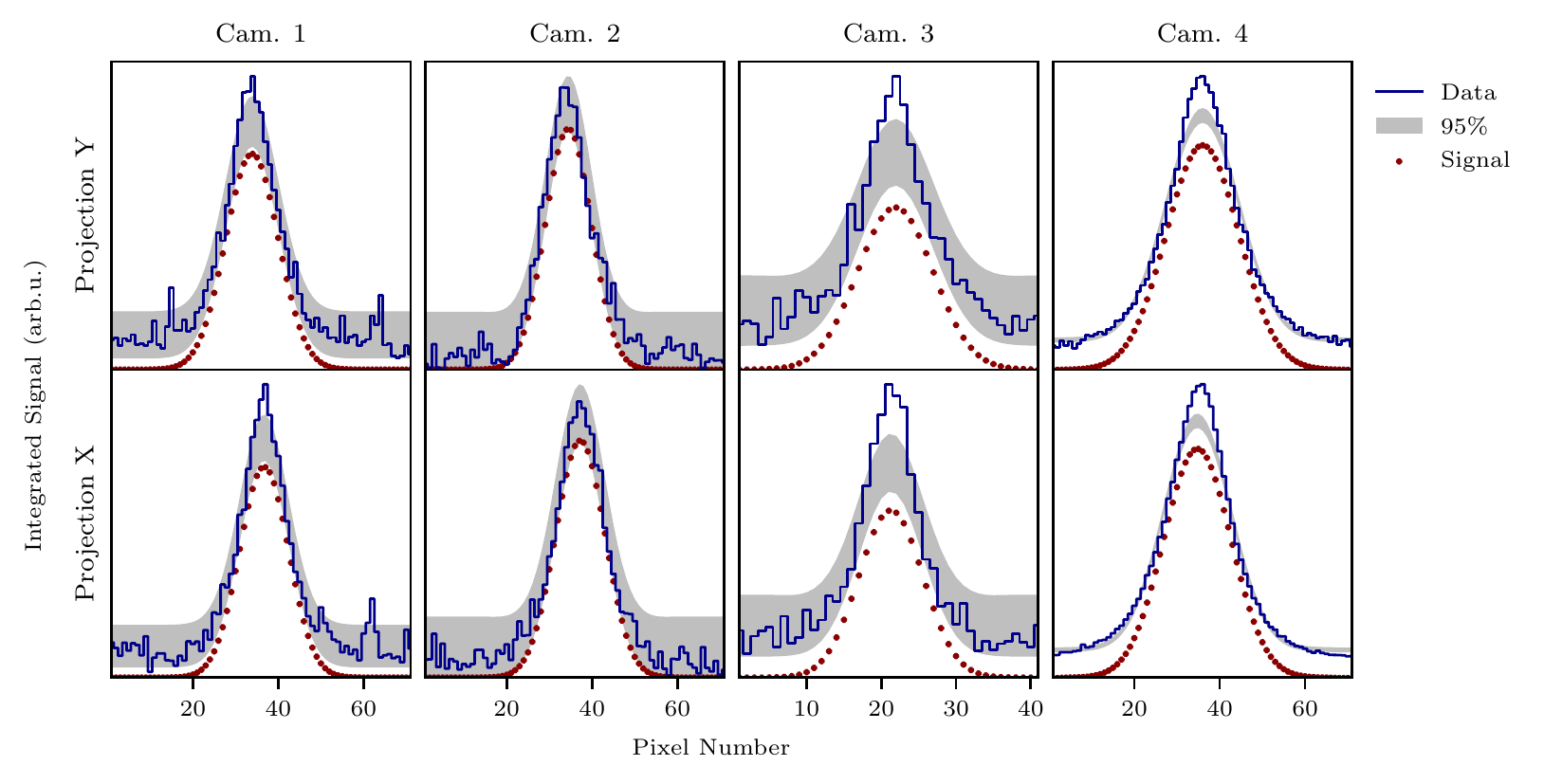}}
\\
\subfloat{\includegraphics{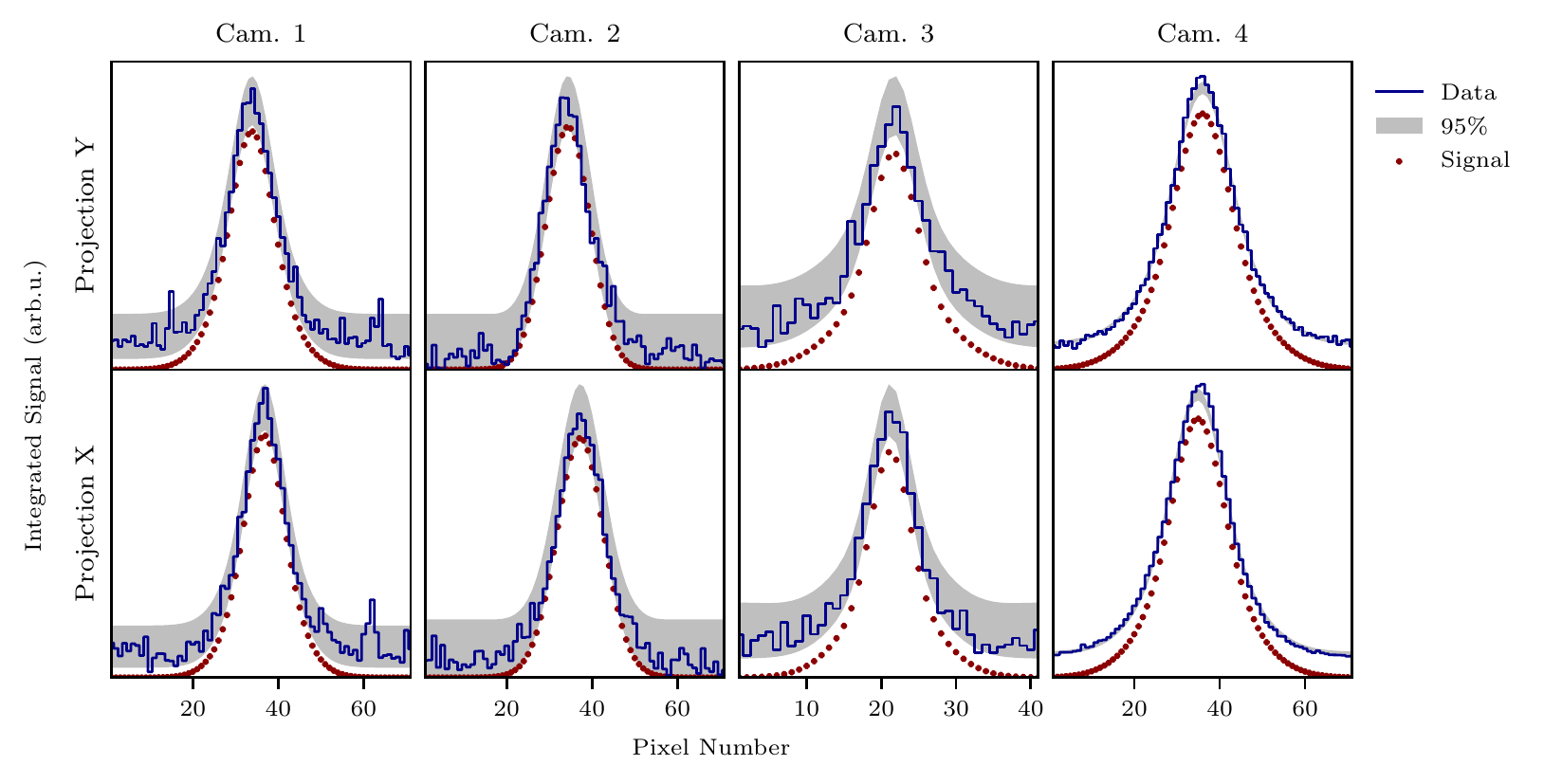}}
\caption{A comparison of the experimental data and the best-fit result for the single (top) and double (bottom) Gaussian models. The blue step-lines show data from one event integrated over rows/columns. The model prediction is plotted for the mean value of the posterior. The grey-filled regions show the 95\% central probability intervals of the model including background and signal. The grey band for Cam.~4 overlaps with the data and is not visible. }
\label{fig:m12-proj}
\end{figure*}

\begin{figure}[!t]
\includegraphics{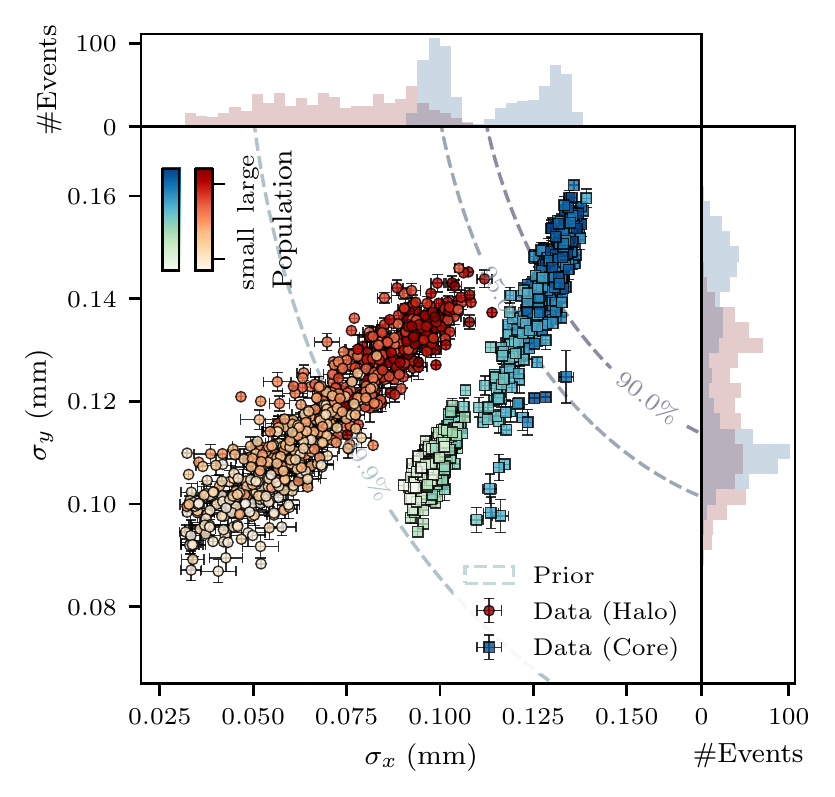}
\caption{The transverse size of the proton bunch at the waist position is shown for events with different bunch populations using a double Gaussian model.  Each event is represented by two symbols that show the halo and core components. Darker colors correspond to a larger bunch population. The color scale is non-linear. Dashed lines represent central intervals of the prior probability distributions. }
\label{fig:tr-size-m-2}
\end{figure}

\begin{figure}[t]
\includegraphics{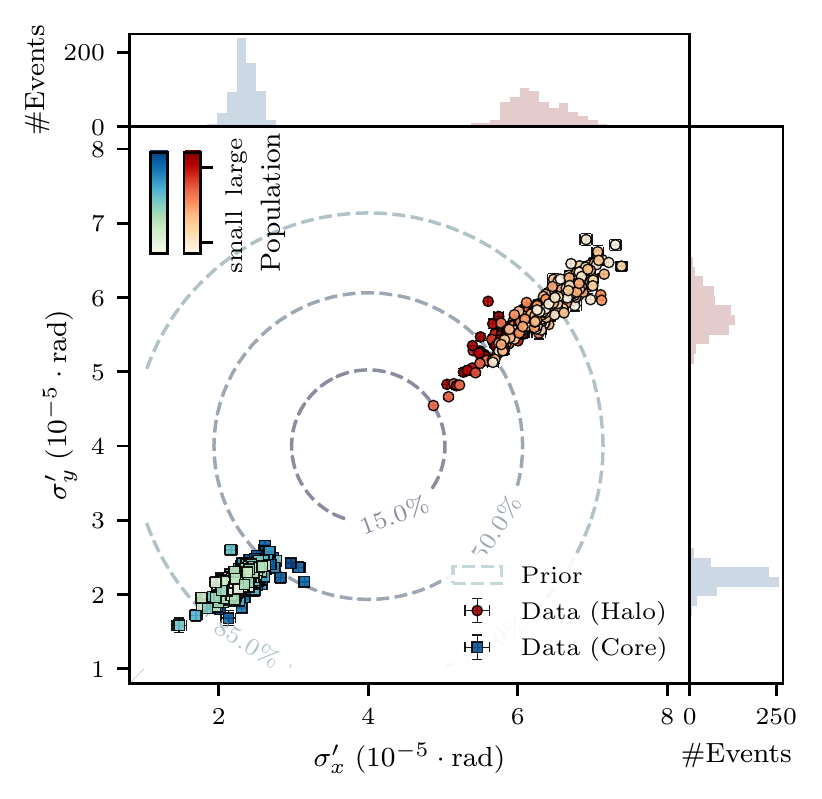}
\caption{The x and y components of the angular spread of the proton bunch are shown for different bunch populations using a double Gaussian model.  Each event is represented by two symbols that show the halo and core components. Darker colors correspond to a larger bunch population. The color scale is non-linear. Dashed lines represent central intervals of the prior probability distributions. }
\label{fig:ang-spread-m-2}
\end{figure}

\begin{figure}[t]
\includegraphics{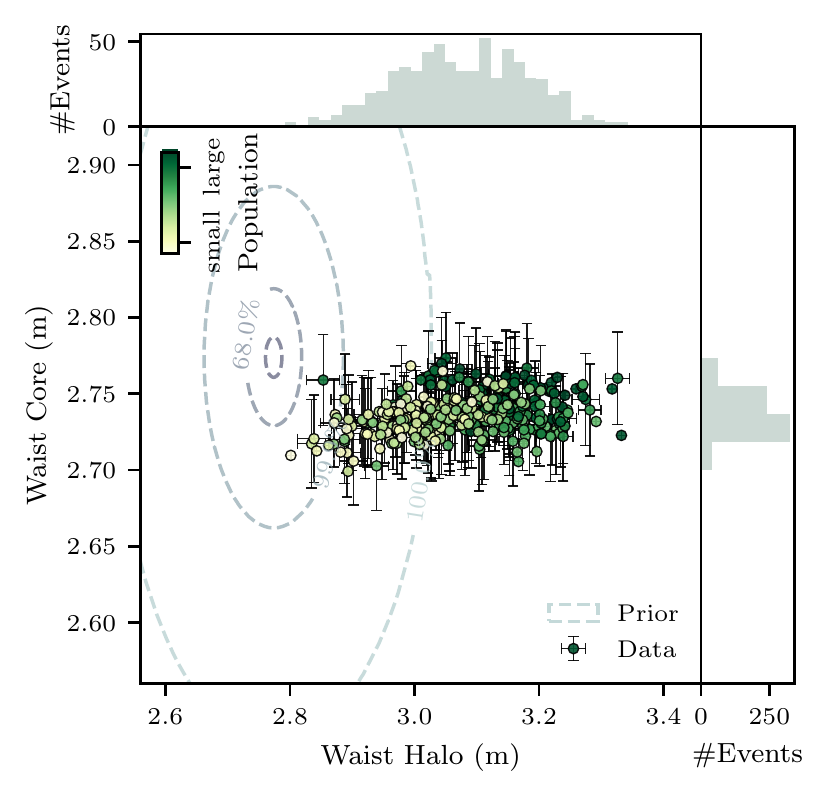}
\caption{The figure shows the means and standard deviations of the waist positions of the halo and core components of the proton bunch. Darker colors correspond to a larger bunch population. The color scale is non-linear. Dashed lines represent central intervals of the prior probability distributions. }
\label{fig:waist-m-2}
\end{figure}

\subsection{Parameters and Priors}

The parameters that describe the proton bunch distribution are
\begin{equation}
\begin{split}
\boldsymbol{\lambda}_{SG} & = \left \{\boldsymbol{\sigma}, \boldsymbol{\sigma}',  z_{w} \right \}, \\ 
\boldsymbol{\lambda}_{DG} & = \left \{\boldsymbol{\sigma}_{c},  \boldsymbol{\sigma}_{h}, \boldsymbol{\sigma}'_{c},  \boldsymbol{\sigma}'_{h}, z_{w, c}, z_{w, h} , \alpha  \right \},
\end{split}
\end{equation} 
where the two vectors correspond to the single and double Gaussian models, respectively. In addition, the nuisance parameters are 
\begin{equation}
\boldsymbol{\nu} = \left \{\boldsymbol{\mu}_j, \Delta x_j, \Delta y_j, \boldsymbol{\tilde{\sigma}}_j, \boldsymbol{i}_j, \boldsymbol{f}_j, p_4 \right \},
\end{equation} 
where $j = 1 \;..\; 4$ denotes the camera index, and the bold font is used for the parameters that have the $x$ and $y$ components. A summary of all the parameters is given in Table~\ref{tab:posterior-summary}. 

The prior probability distributions for the proton bunch parameters are selected based on the AWAKE design report. The priors of the transverse size of the core and halo components of the bunch at the waist position (denoted as $\sigma_{c, x}, \sigma_{c, y}, \sigma_{h, x}, \sigma_{h, y}$) are described by Gaussian probability distributions with means of $\SI{0.2}{\milli\meter}$ and standard deviations of $\SI{0.04}{\milli\meter}$. 
The priors of the angular divergences of the bunch (denoted as $\sigma'_{c, x}, \sigma'_{c, y}, \sigma'_{h, x}, \sigma'_{h, y}$) are described by Gaussian probability distributions with means of $4\times10^{-5}~\si{\radian}$ and standard deviations of $2\times10^{-5}~\si{\radian}$. 
Initially, very broad prior ranges were considered for the proton bunch parameters. After learning about the typical locations of the posteriors, the prior ranges were restricted to reduce computational time.  We truncate the angular divergence of the core component, denoted as $\sigma'_{c, x}, \sigma'_{c, y}$, to the range $[1\times10^{-5}, 8\times10^{-5}]~\si{\radian}$; and the halo component, denoted as $\sigma'_{h, x}, \sigma'_{h, y}$, to the range $[1\times10^{-5}, 4\times10^{-5}]~\si{\radian}$ to clearly identify the halo and core components. The prior probability distributions for the waist positions of the core and halo components, denoted as $z_{w, c}, z_{w, h} $,  are described by a Gaussian distribution with means of $\SI{2.774}{\meter}$ and standard deviations of $\SI{0.03}{\meter}$. The coefficient that defines the intensity ratio for the halo and core component is denoted as $\alpha$ and is described by a uniform prior. 

Parameters that represent the bunch centroid at the camera $j$ along the $x$ and $y$ directions are denoted as $\mu_{x, j}, \mu_{y, j}$, and they are given by uniform prior probability distributions in the ranges specified in Table~\ref{tab:posterior-summary}. The pixel sizes along the $x$ and $y$ directions are denoted as $\Delta x_j$ and $\Delta y_j$, and they are represented by the Dirac delta prior distributions (further denoted as 'constant prior'). The resolution parameters along the $x$ and $y$ directions, denoted as $\tilde{\sigma}_{j,x}, \tilde{\sigma}_{j,y}$, are assumed to be constant for the cameras that have OTR screens. The prior for the camera with the scintillating screen was modeled with a mean of $3$ pixels and a standard deviation of $1.5$ pixels. The conversion of the proton bunch distribution into the pixel signal is performed by defining proportionality coefficients $i_j$, that are described by uniform priors. The standard deviations of the Gaussian fluctuations of the light are defined by $f_j$ (see section~\ref{sec:4}). This parameter has a constant prior for those cameras where the numerical convolutions of the background and signal are computed. For the last camera, the signal fluctuation ($f_4$) and the mean of the background ($p_4$) are kept as free variables with uniform priors. 

\subsection{Results}

We analyzed each event summarized in Table~\ref{table:datasets} independently, using single and double Gaussian bunch models. A comparison of the data from one event (with a small bunch population and the bunch rotation OFF) to the best-fit predictions from the two models is shown in Fig.~\ref{fig:m12-proj}. It can be seen that the double Gaussian model fits the data more closely than the single Gaussian model. The prediction from the single Gaussian model shows a visible discrepancy with the data in the first, third, and fourth cameras. It is especially evident for the events with a small bunch population. A much better data-model agreement is reached for the double Gaussian model, showing that the data fluctuations are covered reliably by the 95\% central interval of the posterior probability distribution. A small disagreement of the distribution for Cam.~4 (see the bottom subplot in Fig.~\ref{fig:m12-proj}) resulted from the fact that some of the pixels were discarded from the analysis due to their saturated signals. In the following, we discuss only the results obtained by using the double Gaussian proton bunch model. 

The transverse sizes of the halo and core components of the bunch at the waist position are shown in Fig.~\ref{fig:tr-size-m-2}. The size of the core is larger than the size of the halo for all events, and both components are significantly smaller than the mode of the prior distribution. There is a correlation of the bunch size with increasing bunch population. The transverse bunch profile is elliptical. 

The angular divergences for different events are shown in Fig.~\ref{fig:ang-spread-m-2}. Events with a larger bunch population have a smaller angular divergence of the halo component. 

The waist positions of the halo and core components are shown in Fig.~\ref{fig:waist-m-2}. The figure shows that the core component of the bunch is focused much closer to the expected waist position compared to the halo component, which is focused further downstream. 

\begin{figure}[!t]
\includegraphics{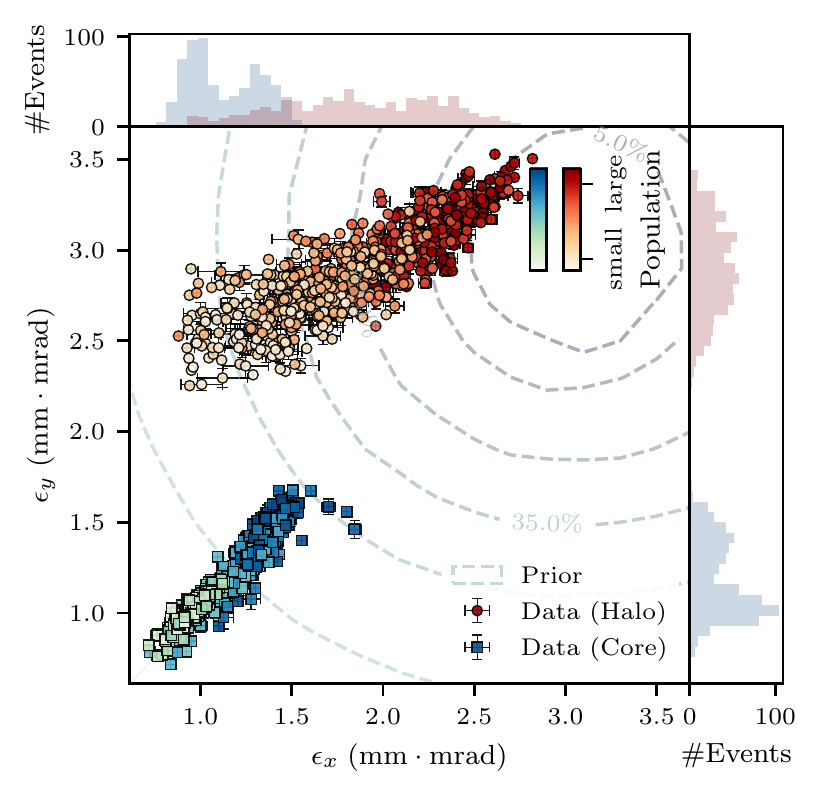}
\caption{The x and y components of the bunch emittance are shown for different bunch populations using a double Gaussian model. Each event is represented by two symbols that indicate the halo and core components. Darker colors correspond to a larger bunch population. The color scale is non-linear. Dashed lines represent central intervals of the prior probability distributions. }
\label{fig:em-m-2}
\end{figure}

The bunch size at the waist and angular divergence can be used to compute the bunch emittance using the following engineering formula
\begin{equation}
\label{eq:emm}
\epsilon = 426.0\cdot10^3 \cdot \sigma [\si{\milli\meter}] \cdot \sigma' [10^{-5}~\si{\radian}].
\end{equation} 
The posterior distributions of the halo and core emittances are summarized in Fig.~\ref{fig:em-m-2}. It can be seen that the bunch emittance increases with the increasing bunch population for both components. The x component of the emittance for the halo component is more strongly correlated with the bunch population than the y component. 

\begin{figure}[t]
    \centering
    \includegraphics{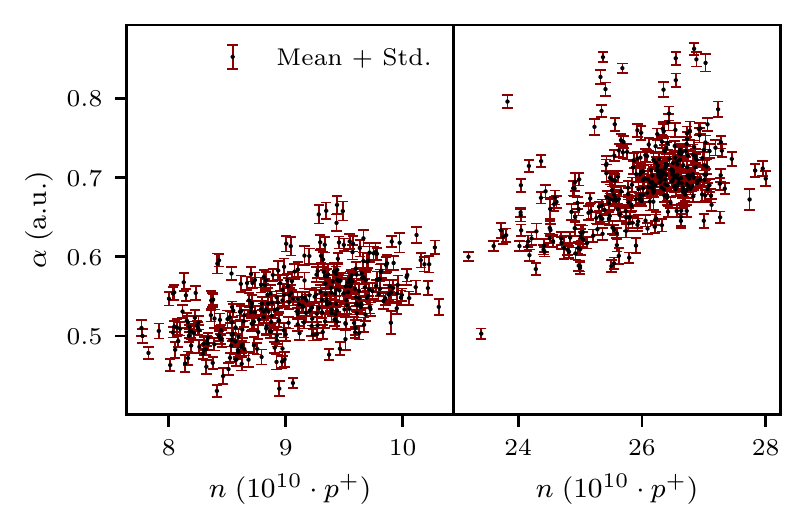}
    \caption{Parameter $\alpha$ that defines the significance of the halo component is shown versus the bunch population. A more significant halo component is observed in the bunches with a larger bunch population.}
    \label{fig:alpha-par}
\end{figure}

The coefficient that shows the intensity of halo and core components is shown in Fig.~\ref{fig:alpha-par}. There is a correlation of $\alpha$ with the proton bunch population that shows that the halo component is more significant for the events with a larger bunch population.

\begin{figure}[t]
    \centering
    \includegraphics{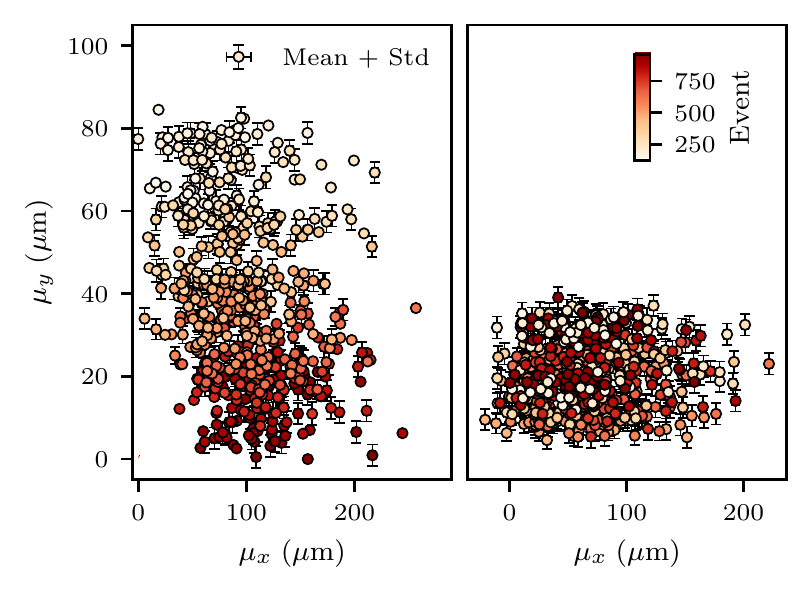}
    \caption{Positions of the bunch centroids at the waist. The left plot shows the drift of the bunch centroid with the time of the measurement. The right plot shows the same data with subtracted time drift.}
    \label{fig:drift}
\end{figure}

In the analysis, we use nuisance parameters to determine the coordinates of the bunch at each screen. These coordinates can be combined to determine the relative alignments of the screens together with polar and azimuthal angles of the individual bunch centroids. By propagating individual bunch centroids to the waist position, the drift and jitter of the bunch can be computed.  Fig.~\ref{fig:drift} shows that the bunch center drifts as a function of event number, which is proportional to the time at which measurement occurred. The standard deviation at the waist is $\sigma(\mu_x) \approx \SI{43}{\micro\meter}$, $ \sigma(\mu_y) \approx \SI{20}{\micro\meter}$. Once the time-drift of the bunch is subtracted, the resulting jitters are $\sigma(\mu_x) \approx \SI{41}{\micro\meter}$ and $ \sigma(\mu_y) \approx \SI{8}{\micro\meter}$. 

A summary of the average posterior parameters for events with small and large bunch populations is given in Table~\ref{tab:posterior-summary} for the double Gaussian model. A summary of the average proton bunch parameters for the single Gaussian model is given in Table~\ref{tab:posterior-summary-sg}. 

\begin{table*}
\caption{The table summarizes the parameters used in the analysis of the proton bunch with the double Gaussian model. Parameters are separated into the proton bunch ($\boldsymbol{\lambda}$), nuisance ($\boldsymbol{\nu}$), and calculated categories. The fourth column describes prior probability distributions.  If the distribution is uniform on a certain region or truncated, the corresponding region is specified in rectangular parentheses; a single number denotes the argument of the Dirac delta distribution, $\mathcal{N}(\mu, \sigma)$ stands for a Gaussian distribution with a mean $\mu$ and a standard deviation $\sigma$. The fifth and sixth columns show the mean and standard deviation of the parameters averaged over the datasets with small (1) and large (2) bunch populations. }
\begin{ruledtabular}
\begin{tabular}{lccccc}
\textbf{Parameter}& \textbf{Symbol} & \textbf{Unit} & \textbf{Prior} & \textbf{Posterior \; 1} & \textbf{Posterior \; 2}\\ \hline
\textit{Proton Bunch}  \\ \hline
Transverse size x, core & $\sigma_{c, x}$ &\si{\milli\meter} & $\mathcal{N}(0.2, 0.04) $ & $0.099 \pm 0.0031$  & $0.13 \pm 0.0065$\\
Transverse size y, core & $\sigma_{c, y}$ &\si{\milli\meter} & $\mathcal{N}(0.2, 0.04) $ & $0.11 \pm 0.0041$  & $0.14\pm 0.012$ \\ 
Transverse size x, halo & $\sigma_{h, x}$ &\si{\milli\meter} & $\mathcal{N}(0.2, 0.04) $ & $0.056 \pm 0.012$  & $0.086\pm 0.011$ \\ 
Transverse size y, halo & $\sigma_{h, y}$ &\si{\milli\meter} & $\mathcal{N}(0.2, 0.04) $ & $0.11 \pm 0.0079$  & $0.13 \pm 0.0069$  \\
Angular spread x, core  & $\sigma'_{c, x}$ &\SI{d-5}{\radian} & $\mathcal{N}(4.0, 2.0) $ & $2.28 \pm 0.14$   &   $2.41 \pm 0.18$  \\  
Angular spread y, core  & $\sigma'_{c, y}$ &\SI{d-5}{\radian} & $\mathcal{N}(4.0, 2.0) $ & $2.19 \pm 0.11$   &   $2.25 \pm  0.15$  \\ 
Angular spread x, halo  & $\sigma'_{h, x}$ &\SI{d-5}{\radian} & $\mathcal{N}(4.0, 2.0) $ & $6.5 \pm 0.31$  &  $5.99 \pm 0.26$ \\  
Angular spread y, halo  & $\sigma'_{h, x}$ &\SI{d-5}{\radian} & $\mathcal{N}(4.0, 2.0) $ & $5.95 \pm 0.27$   &   $5.6 \pm 0.22$  \\ 
Waist position, core &  $z_{w, c}$ &\si{\meter}  & $\mathcal{N}(2.774, 0.03)$ & $2.73 \pm 0.011$   &   $2.74 \pm 0.013$ \\ 
Waist position, halo &  $z_{w, h}$ &\si{\meter} & $\mathcal{N}(2.774, 0.03)$ & $3.01 \pm 0.082$   &   $3.14 \pm 0.081$ \\
Intensity ratio &  $\alpha$ & one & $[0.25,1.0]$ & $0.54 \pm 0.041$   &   $0.69 \pm 0.05$     \\  \hline

\textit{Nuisance} \\  \hline

Alignment on Cam. 1, x & $\mu_{1, x}$ & px & $[23.0,48.0]$ & $35.2 \pm 1.79$  & $ 35.6 \pm 2.06 $\\
Alignment on Cam. 2, x & $\mu_{2, x}$ & px & $[23.0,48.0]$ & $36.2 \pm 2.06$  & $ 36.7\pm 2.38 $\\
Alignment on Cam. 3, x & $\mu_{3, x}$ & px & $[10.0,30.0]$ & $21.4 \pm 0.362 $ & $21.5 \pm 0.373$\\
Alignment on Cam. 4, x & $\mu_{4, x}$ & px & $[23.0,48.0]$ & $35.7 \pm 0.508$  & $35.3 \pm 0.45$\\
Alignment on Cam. 1, y & $\mu_{1, y}$ & px & $[23.0,48.0]$ & $35.4 \pm 0.484$  & $ 35.3\pm 1.32 $\\
Alignment on Cam. 2, y & $\mu_{2, y}$ & px & $[23.0,48.0]$ & $36.0 \pm 0.483$  & $ 36.0 \pm 1.22 $\\
Alignment on Cam. 3, y & $\mu_{3, y}$ & px & $[10.0,30.0]$ & $20.6 \pm 0.275$  & $20.7 \pm 0.356$\\
Alignment on Cam. 4, y & $\mu_{4, y}$ & px & $[23.0,48.0]$ & $34.6 \pm 0.201$  & $34.7 \pm 0.575$\\
Pixel size on Cam. 1, x & $\Delta x_1$  & \si{\micro\meter}  & $27.1$ & -  & -\\
Pixel size on Cam. 2, x & $\Delta x_2$  & \si{\micro\meter}  & $21.6$ & -  & -\\
Pixel size on Cam. 3, x & $\Delta x_3$  & \si{\micro\meter}  & $114.0$ & -  & -\\
Pixel size on Cam. 4, x & $\Delta x_4$  & \si{\micro\meter}  & $121.8$ & -  & -\\
Pixel size on Cam. 1, y & $\Delta y_1$  & \si{\micro\meter}  & $30.5$ & -  & -\\
Pixel size on Cam. 2, y & $\Delta y_2$  & \si{\micro\meter}  & $23.4$ & -  & -\\
Pixel size on Cam. 3, y & $\Delta y_3$  & \si{\micro\meter}  & $125.0$ & -  & -\\
Pixel size on Cam. 4, y & $\Delta y_4$  & \si{\micro\meter}  & $120.0$ & -  & -\\
Resolution effect on Cam. 1, x & $\tilde{\sigma}_{1,x}$  & px  & $1.0$ & -  & -\\
Resolution effect on Cam. 2, x & $\tilde{\sigma}_{2,x}$  & px  & $1.0$ & -  & -\\
Resolution effect on Cam. 3, x & $\tilde{\sigma}_{3,x}$  & px  & $1.0$ & -  & -\\
Resolution effect on Cam. 4, x & $\tilde{\sigma}_{4,x}$  & px  & $\mathcal{N}(3.0, 1.5)$ & $4.6 \pm 0.15$  & $4.7 \pm 0.19$ \\
Resolution effect on Cam. 1, y & $\tilde{\sigma}_{1,y}$  & px  & $1.0$ & -  & -\\
Resolution effect on Cam. 2, y & $\tilde{\sigma}_{2,y}$  & px  & $1.0$ & -  & -\\
Resolution effect on Cam. 3, y & $\tilde{\sigma}_{3,y}$  & px  & $1.0$ & -  & -\\
Resolution effect on Cam. 4, y & $\tilde{\sigma}_{4,y}$  & px  & $\mathcal{N}(3.0, 1.5)$ & $4.1 \pm 0.13$  & $4.5 \pm 0.2$ \\
Signal amplitude on Cam 1 & $i_1$ & counts & $[1.0, 13.0]$ & $3.0 \pm 0.185$  & $8.31 \pm 0.329$\\
Signal amplitude on Cam 2 & $i_2$ & counts & $[1.0, 17.0]$ & $3.9 \pm 0.241$ & $11.1 \pm 0.381$\\
Signal amplitude on Cam 3 & $i_3$ & counts & $[1.0, 5.0]$ & $2.51\pm 0.168$  & $2.67 \pm 0.117$\\
Signal amplitude on Cam 4 & $i_4$ & counts & $[1.0, 13.0]$ & $2.8 \pm 0.17$  & $8.59 \pm 0.303$\\
Signal fluctuations on Cam 1 & $f_1$ & one & $2.0$ & -  & -\\
Signal fluctuations on Cam 2 & $f_2$ & one & $2.0$ & -  & -\\
Signal fluctuations on Cam 3 & $f_3$ & one & $2.0$ & -  & -\\
Signal fluctuations on Cam 4 & $f_4$ & one & $[1.0, 3.0]$ & $1.62 \pm 0.22$  & $1.78 \pm 0.414$\\
Pedestal on Cam 4 & $p_4$ & counts & $[4.0, 40.0]$ & $18.4 \pm 1.04$  & $27.7\pm 2.97$\\ \hline
\textit{Calculated} \\  \hline
Emittance x, core &  $\epsilon_{c, x}$ &\si{\milli\meter\milli\radian} & - & $0.96 \pm 0.07$   &   $1.3 \pm 0.13$ \\
Emittance y, core &  $\epsilon_{c, y}$ &\si{\milli\meter\milli\radian}  & - & $1.02 \pm 0.074$ &  $1.4 \pm 0.16$\\ 
Emittance x, halo &  $\epsilon_{h, x}$ &\si{\milli\meter\milli\radian} & - & $1.5 \pm 0.3$   &   $2.2 \pm 0.27$ \\
Emittance y, halo &  $\epsilon_{h, y}$ &\si{\milli\meter\milli\radian}  & - & $2.7 \pm 0.18$  &  $3.1 \pm 0.17$ \\ 
Bunch population &  $n$ & $10^{10} p^+$  & - & $9.1 \pm 0.6$  &  $25.9 \pm 0.9$  \\ 
\end{tabular}
\end{ruledtabular}
\label{tab:posterior-summary}
\end{table*}

\begin{table*}
\caption{The table summarizes the proton bunch parameters using the single Gaussian bunch model. The fourth column describes prior probability distributions. The fifth and sixth columns show the mean and standard deviation of the parameters averaged over the datasets with small (1) and large (2) bunch populations.}
\begin{ruledtabular}
\begin{tabular}{lccccc}
\textbf{Parameter}& \textbf{Symbol} & \textbf{Unit} & \textbf{Prior} & \textbf{Posterior \; 1} & \textbf{Posterior \; 2}\\ \hline
\textit{Proton Bunch}  \\ \hline
Transverse size x & $\sigma_{x}$ &\si{\milli\meter} & $\mathcal{N}(0.20, 0.04) $ & $0.089 \pm 0.004$  & $0.11 \pm 0.006$ \\
Transverse size y & $\sigma_{y}$ &\si{\milli\meter} & $\mathcal{N}(0.20, 0.04) $ & $0.11 \pm 0.004$  & $0.14 \pm 0.006$ \\ 
Angular spread x  & $\sigma'_{x}$ &\SI{d-5}{\radian} & $\mathcal{N}(4.0, 2.0) $ & $4.42\pm 0.24$    &   $4.73 \pm 0.17$   \\  
Angular spread y  & $\sigma'_{y}$ &\SI{d-5}{\radian} & $\mathcal{N}(4.0, 2.0) $ & $4.14 \pm 0.24$   &   $4.43 \pm 0.17$  \\ 
Waist position &  $z_{w}$ &\si{\meter}  & $\mathcal{N}(2.774, 0.03)$ & $2.98 \pm 0.046$   &   $3.2 \pm 0.055$ \\ 
\hline
\textit{Calculated} \\  \hline
Emittance x &  $\epsilon_{x}$ &\si{\milli\meter\milli\radian} & - & $1.7\pm 0.12$   &   $2.1 \pm 0.16$ \\
Emittance y &  $\epsilon_{y}$ &\si{\milli\meter\milli\radian} & - & $2.0 \pm 0.14$   &   $2.6 \pm 0.17$ \\
\end{tabular}
\end{ruledtabular}
\label{tab:posterior-summary-sg}
\end{table*}

\section{Conclusions}
\label{sec:6}

We have developed and applied an approach to analyzing the parameters of the proton bunch in the AWAKE experiment. In this approach, the data from multiple beam imaging systems that capture integrated radial bunch profiles were used to determine optimal parameters of the model that describe proton bunch propagation along the beamline. The fitting procedure was performed using a Bayesian approach, and the MCMC sampling was used to extract the posterior distributions. This approach will be used in future AWAKE runs to analyze the stability of the proton bunch parameters over long runs. 

Two models that describe the radial bunch density were considered. In the first model, the transverse bunch profile was represented as a Gaussian function, in the second model --- as a mixture of two Gaussians denoted as halo and core. By definition, the halo component has a larger emittance compared to the core. These models have been tested using simulated events, and the results show that reconstruction of true parameters is possible with typical uncertainties of a few percent. 

We have acquired a dataset with 671 proton bunch extractions, each characterized by varying proton bunch parameters, and applied the developed analysis scheme to this experimental data. It has been demonstrated that the double Gaussian model gives much better agreement with the experimental data compared to the single Gaussian model. The resulting posterior parameters for the double Gaussian model indicate the following:

\begin{itemize}
    \item[\textendash] The transverse size of the proton bunch at the waist position is smaller than the baseline parameters for all bunch populations. 
    \item[\textendash]  Sizes of the halo and core components increase with the increasing bunch population.
    \item[\textendash]  The transverse bunch profile is elliptical (the horizontal size is smaller than the vertical by $\approx8\%$ and $\approx40\%$ for the core and halo components, respectively).  
    \item[\textendash] The contribution of the halo component increases with the bunch population.
    \item[\textendash] The bunch emittance is smaller than the nominal parameters. 
\end{itemize}

A systematic drift of the bunch centroid was observed of approximately $\SI{50}{\micro\meter}$ during 6 hours of measurements. The jitter of the bunch at the waist position after drift correction is $\sigma(\mu_x) \approx \SI{41}{\micro\meter}$ and $ \sigma(\mu_y) \approx \SI{8}{\micro\meter}$.

\begin{acknowledgments}

We acknowledge useful discussions of the presented analysis approach and results with the SPS team. 
This work was supported in parts by a Leverhulme Trust Research Project Grant RPG-2017-143 and by STFC (AWAKE-UK, Cockcroft Institute core, John Adams Institute core, and UCL consolidated grants), United Kingdom;
a Deutsche Forschungsgemeinschaft project grant PU 213-6/1 ``Three-dimensional quasi-static simulations of beam self-modulation for plasma wakefield acceleration'';
the National Research Foundation of Korea (Nos.\ NRF-2016R1A5A1013277 and NRF-2020R1A2C1010835);
the Portuguese FCT---Foundation for Science and Technology, through grants CERN/FIS-TEC/0032/2017, PTDC-FIS-PLA-2940-2014, UID/FIS/50010/2013 and SFRH/IF/01635/2015;
the U.S.\ National Science Foundation under grant PHY-1903316;
the Wolfgang Gentner Programme of the German Federal Ministry of Education and Research (grant no.\ 05E15CHA);
M. Wing acknowledges the support of DESY, Hamburg.
Support of the National Office for Research, Development and Innovation (NKFIH) under contract number 2019-2.1.6-NEMZ\_KI-2019-00004 and the support of the Wigner Datacenter Cloud facility through the Awakelaser project is acknowledged.
The work of V. Hafych has been supported by the European Union's Framework Programme for Research and Innovation Horizon 2020 (2014--2020) under the Marie Sklodowska-Curie Grant Agreement No.\ 765710.
The AWAKE collaboration acknowledge the SPS team for their excellent proton delivery.

\end{acknowledgments}

\bibliography{apssamp}

\end{document}